\documentclass[twocolumn]{aastex63}

\usepackage{hyperref}
\newcommand{\uat}[2]{\href{http://astrothesaurus.org/uat/#2}{#1 (#2)}}
\usepackage{natbib}
\usepackage{amsmath}
\usepackage{chngcntr}
\usepackage{CJK}
\received{15th Dec. 2022}
\revised{14th Feb. 2023}
\accepted{27th Mar. 2023}
\submitjournal{ApJL}

\shorttitle{BTFR of HUDGs}
\shortauthors{H. Hu et al.}

\begin{document}
\begin{CJK*}{UTF8}{gbsn}
\title{Global dynamic scaling relations of H{\sc i}-rich ultra-diffuse galaxies}

\correspondingauthor{Qi Guo, Hui-Jie Hu}
\email{guoqi@nao.cas.cn, huhuijienao@gmail.com}

\author[0000-0002-1908-0384]{Hui-Jie Hu}
\affiliation{National Astronomical Observatories, Chinese Academy of Science, 20A Datun Road, Beijing 100101, China}
\affiliation{University of Chinese Academy of Sciences, No. 19 A Yuquan Road, Beijing 100049, China}

\author[0000-0002-7972-3310]{Qi Guo}
\affiliation{National Astronomical Observatories, Chinese Academy of Science, 20A Datun Road, Beijing 100101, China}
\affiliation{University of Chinese Academy of Sciences, No. 19 A Yuquan Road, Beijing 100049, China}
\affiliation{Institute for Frontiers in Astronomy and Astrophysics, Beijing Normal University, Beijing 102206, China}

\author{Zheng Zheng}
\affiliation{National Astronomical Observatories, Chinese Academy of Science, 20A Datun Road, Beijing 100101, China}
\affiliation{Research Center for Intelligent Computing Platforms, Zhejiang Laboratory, Hangzhou 311100, China}

\author[0000-0003-3279-0134]{Hang Yang}
\affiliation{National Astronomical Observatories, Chinese Academy of Science, 20A Datun Road, Beijing 100101, China}
\affiliation{University of Chinese Academy of Sciences, No. 19 A Yuquan Road, Beijing 100049, China}

\author[0000-0002-9390-9672]{Chao-Wei Tsai}
\affiliation{National Astronomical Observatories, Chinese Academy of Science, 20A Datun Road, Beijing 100101, China}
\affiliation{Institute for Frontiers in Astronomy and Astrophysics, Beijing Normal University, Beijing 102206, China} 
\affiliation{University of Chinese Academy of Sciences, No. 19 A Yuquan Road, Beijing 100049, China}

\author[0000-0003-1632-2541]{Hong-Xin Zhang}
\affiliation{Key Laboratory for Research in Galaxies and Cosmology, Department of Astronomy, University of Science and Technology of
China, Hefei 230026, China}
\affiliation{School of Astronomy and Space Science, University of Science and Technology of China, Hefei 230026, China}

\author[0000-0002-7299-2876]{Zhi-Yu Zhang}
\affiliation{School of Astronomy and Space Science, Nanjing University, Nanjing 210023, People’s Republic of China}
\affiliation{Key Laboratory of Modern Astronomy and Astrophysics (Nanjing University), Ministry of Education, Nanjing 210023, People’s Republic of China}



\begin{abstract}
The baryonic Tully-Fisher relation (BTFR), which connects the baryonic mass of galaxies with their circular velocities, has been validated across a wide range of galaxies, from dwarf galaxies to massive galaxies. Recent studies have found that several ultra-diffuse galaxies (UDGs) deviate significantly from the BTFR, indicating a galaxy population with abnormal dynamical properties. However, such studies were still confined within a small sample size. In this study, we used the 100\% complete Arecibo Legacy Fast Arecibo L-band Feed Array (ALFALFA) to investigate the BTFR of 88 H{\sc i}-rich UDGs (HUDGs), which is the largest UDG sample with dynamical information. We found that the HUDGs form a continuous distribution in the BTFR diagram, with high-velocity galaxies consistent with normal dwarf galaxies at 1 $\sigma$ level, and low-velocity galaxies deviating from the BTFR, in line with that reported in the literature. We point out that the observed deviation may be subject to various selection effects or systemic biases. Nevertheless, we found that the significance of the deviation of HUDGs from the BTFR and TFR are different, i.e., they either deviate from the BTFR or from the TFR. Our result indicates that a high-gas fraction may play an important role in explaining the deviation of HUDGs from BTFR. 

\end{abstract}



\section{Introduction}

The Tully-Fisher relation \citep[TFR;][]{1977A&A....54..661T}, which correlates luminosity and circular velocity of spiral galaxies, serves as a standard ruler for the measurement of cosmic distance. However, it varies with galaxy properties, including colors and ages \citep{2001ApJ...550..212B,2003ApJS..149..289B}. \cite{1999ASPC..170....3F} and \cite{2000ApJ...533L..99M} found that the baryonic Tully-Fisher relation (BTFR) could be a more fundamental relation than TFR. This power law relation between the baryonic mass and circular velocity \citep[e.g.][]{2008MNRAS.386..138B,2012AJ....143...40M,2016ApJ...816L..14L,2017AJ....153....6K} is even tighter than the TFR. The BTFR spans a wider mass range (six orders in baryonic mass), including dwarf galaxies and low surface brightness galaxies. 

Theoretically, both TFR and BTFR originate from the virial theorem, modulated by multiple physical processes. Thus they could pose constrains on galaxy formation models. Modern simulations \citep[e.g.][]{2011ApJ...742...16T,2017MNRAS.464.2419S} predict a break-point at $10^9M_{\odot}$, below which BTFR bends down towards low masses, in conflict with observations, they discussed that this might be caused by observational biases. However, observations of the Local Volume dwarf galaxies show that dwarf galaxies also follow the BTFR determined by massive galaxies \citep{2008MNRAS.386..138B,2017AJ....153....6K}. 

Ultra-diffuse galaxies (UDGs) constitute a special population of dwarf galaxies, drawing great attention in recent years. They have stellar masses similar to normal dwarf galaxies ($\lesssim 10^9M_{\odot}$) yet with sizes comparable to $L^{\star}$ (the Milky Way analogs) galaxies \citep{2015ApJ...798L..45V,2015ApJ...804L..26V}. The origin of UDGs is still poorly understood. Theoretical studies suggest field UDGs could stem from stellar feedback \citep{2016ApJ...820..131E}, early mergers \citep{2021MNRAS.502.5370W}, and high spins \citep{2017MNRAS.470.4231R,2019MNRAS.490.5182L}, while satellite UDGs could be the descendants of field UDGs and/or dwarf galaxies reshaped by tidal heating \citep{2019MNRAS.487.5272J, 2019MNRAS.490.5182L}.

Recent studies show that UDGs in the BTFR diagram could be different from normal dwarf galaxies. \cite{2018Natur.555..629V,2019ApJ...874L...5V} reported two satellite UDGs in clusters deficient in dark matter using the dynamics of surrounding globular clusters. \cite{2020ApJ...902...39K} presented nine H{\sc i}-rich UDGs (HUDGs) using the Robert C. Byrd Green Bank Telescope (GBT) and found that most of their HUDGs lie above the BTFR. Since they adopt optical morphology to estimate the inclination, it is possible that the misalignment between optical images and H{\sc i} velocity fields leads to a large uncertainty in the circular velocity. Using the Karl G. Jansky Very Large Array (VLA), \cite{2019ApJ...883L..33M} reported six HUDGs that have much higher baryonic masses compared to normal dwarf galaxies with the same circular velocities. Despite the capability of the interferometer in resolving spacial distribution and velocity fields of H{\sc i}, inclination could still be an issue to affect their conclusions. For example, several UDGs investigated by \cite{2019ApJ...883L..33M} have twisted isophotes which makes the determination of inclination uncertain.

In this study, we use 88 HUDGs selected from the largest blind H{\sc i} survey, the 100\% complete Arecibo Legacy Fast Arecibo L-band Feed Array \citep[ALFALFA;][]{2018ApJ...861...49H} catalogue ($\alpha.100$) to revise these relations. The large sample size allows better statistics which helps reducing the uncertainties. In Section \ref{sec:dataaMe}, we briefly describe the sample selection criteria and the methods to extract physical properties. Our main results are presented in Section \ref{sec:res}. The discussion and summary are presented in Section \ref{sec:dis} and Section \ref{sec:sum}.

\section{Data and Methods} \label{sec:dataaMe}
\subsection{Sample selection} \label{sec:sample}
\defcitealias{2020NatAs...4..246G}{G20}
We start with 254\footnote{Rather than 252 reported in \cite{2019MNRAS.490..566J}, we count the UDGs in their table, and find that there are 254 HUDs-B.} broadly\footnote{Mean surface brightness rather than the classical central surface brightness} selected H{\sc i}-bearing UDG candidates (HUDs-B) from the $\alpha.100$ \citep{2019MNRAS.490..566J}. Here UDGs are defined as galaxies with {\it g}-band effective radius $r_{g,eff}>1.5$ kpc, mean surface brightness within the effective radius $\left \langle \mu(r,r_{eff}) \right \rangle > 24$ mag arcsec$^{-2}$, and {\it r}-band absolute magnitude $M_{r}>-17.6$ \citep{2017ApJ...842..133L}. To avoid galaxies with complicated dynamical structures, we further apply the following criteria. 

First, we visually remove 51 galaxies that have multiple optical counterparts within $6'$ (corresponding to twice of Arecibo beam size). Second, to minimize the inclination correction effect, we discard 114 face-on HUDG candidates with large axis ratios (b/a $>$0.7, corresponding to an inclination angle of $\sim 47^{\circ}$). It has been found that photometries of low surface brightness galaxies provided by the Sloan Digital Sky Survey (SDSS) using their standard pipeline could be off by 0.5 mag due to the over-subtraction of the sky background \citep{2007ApJ...660.1186L}. We thus re-process the SDSS images more carefully \citep{2015AJ....149..199D,2015ApJ...800..120Z,2020NatAs...4..246G}. In practice, we apply multi-Gaussian expansion \citep[MGE;][]{2002MNRAS.333..400C} on the SDSS DR12 mosaics images\footnote{https://dr12.sdss.org/mosaics} and derive the axis ratio after masking out foreground stars with SExtractor \citep{1996A&AS..117..393B}. We find that for our 203 parent HUDGs, $58\%$ (117) of them the axis ratio differences (re-processed HUDGs compared to SDSS pipeline) are greater than 0.1, $26\%$ (53) of them greater than 0.2. 
Last, we remove one suspicious galaxy with $M_g - M_r = 3.78$ after re-analyzing their photometries. Our final HUDG sample contains 88 galaxies. 

\subsection{Method} \label{sec:Method}
In this subsection, we describe the methods to calculate the total baryonic mass (the sum of stellar mass and neutral gas mass) and the circular velocity. 

\subsubsection{Baryonic mass} \label{sec:mass}
We assume that the baryonic mass $M_{bary}$ is dominated by stellar mass $M_{\star}$ and neutral atomic gas mass $M_{gas}$, and neglect all ionized and molecular gas phases. For stellar mass, we use the re-processed SDSS Petrosian magnitudes in {\it g} and {\it r} bands and the distances from $\alpha.100$ \citep{2011AJ....142..170H,2020AJ....160..271D}. In $\alpha.100$, for galaxies with $cz>6000$ km s$^{-1}$ distances are estimated as $d = cz/H_0$ where $H_{0}$=70 km s$^{-1}$ Mpc$^{-1}$ is the Hubble constant, while for galaxies with $cz<6000$ km s$^{-1}$, distances are obtained using the local universe peculiar velocity model \citep{2004ApJ...607L.115M} with a combination of primary distances from the literature and secondary from the TFR. If the distances of HUDGs are estimated mainly using the TFR, it would strongly affect our results. Therefore, we examine the distances of our sample galaxies and found that most of them follow the Hubble flow instead of the TFR (see details in Fig. \ref{fig:dist} in the Appendix). We use the stellar mass-to-light ratio recipe by \cite{2003ApJS..149..289B} to calculate the stellar mass
\begin{equation}
\log_{10}({M/L}) = -0.7 -0.15+1.252 \times (g - r)
\end{equation}
Here we use the Kroupa stellar initial mass function (IMF) 
with coefficients given by \cite{2020AJ....160..122D} for low surface brightness galaxies.

Neutral gas mainly consists of neutral hydrogen and Helium. The total gas mass can be derived by assuming the same Helium fraction from the big bang nucleosynthesis: M$_{gas} = 1.33 \times M_{HI}$ \citep[Y$_{P}^{BBN}=0.247$;][]{2020A&A...641A...6P}. H{\sc i} mass is directly retrieved from $\alpha.100$ catalogue. 

\subsubsection{Circular Velocity}
We follow \citealt{2020NatAs...4..246G} (hereafter \citetalias{2020NatAs...4..246G}) to use $w_{20}$, the width at the $20\%$ level of the peak flux of H{\sc i} spectrum, as the indicator of the circular velocity. It has been proven as a better indicator \citep[][\citetalias{2020NatAs...4..246G}]{2019MNRAS.484.3267L} of the circular velocity than $w_{50}$, the width at $50\%$ level of the peak flux-estimates, which is usually used in the literature. The latter underestimated the circular velocity. However, the $w_{20}$ in $\alpha.100$ catalogue is heavily influenced by the noises (Fig. \ref{fig:w20alf} in Appendix). We follow the procedure in \citetalias{2020NatAs...4..246G} and re-analyze the data to extract $w_{20}$. 
For the non-edge-on HUDGs, we also correct the inclination effect. Since the H{\sc i} images and velocity fields are not available, we use the optical {\it g}-band inclination angles instead. The {\it g}-band axis ratio, b/a, is calculated by applying MGE on the re-processed SDSS images. 

The velocity $V_{HI}$ is then estimated as: 
\begin{equation}
{V_{HI} = \frac{w_{20}}{2sin(i)}}
\end{equation}
\begin{equation}
sin(i)  = \sqrt[]{\frac{1-(\frac{b}{a})^2}{1-q_0^2}}\\
\end{equation}
where {\it i} is the {\it g}-band inclination angle, $q_0$ denotes the thickness of the galaxy, i.e. the axis ratio of a galaxy seen from the edge-on direction. Here we adopt $q_0=0.2$ which is usually used in the literature \citep{2009AJ....138..323T}. 

In addition to the HUDGs, we also include 324 dwarf galaxies from \citetalias{2020NatAs...4..246G} for comparison. We perform the same data analysis for these dwarf galaxies as for the HUDGs. 

\begin{figure*}[ht!]
\plotone{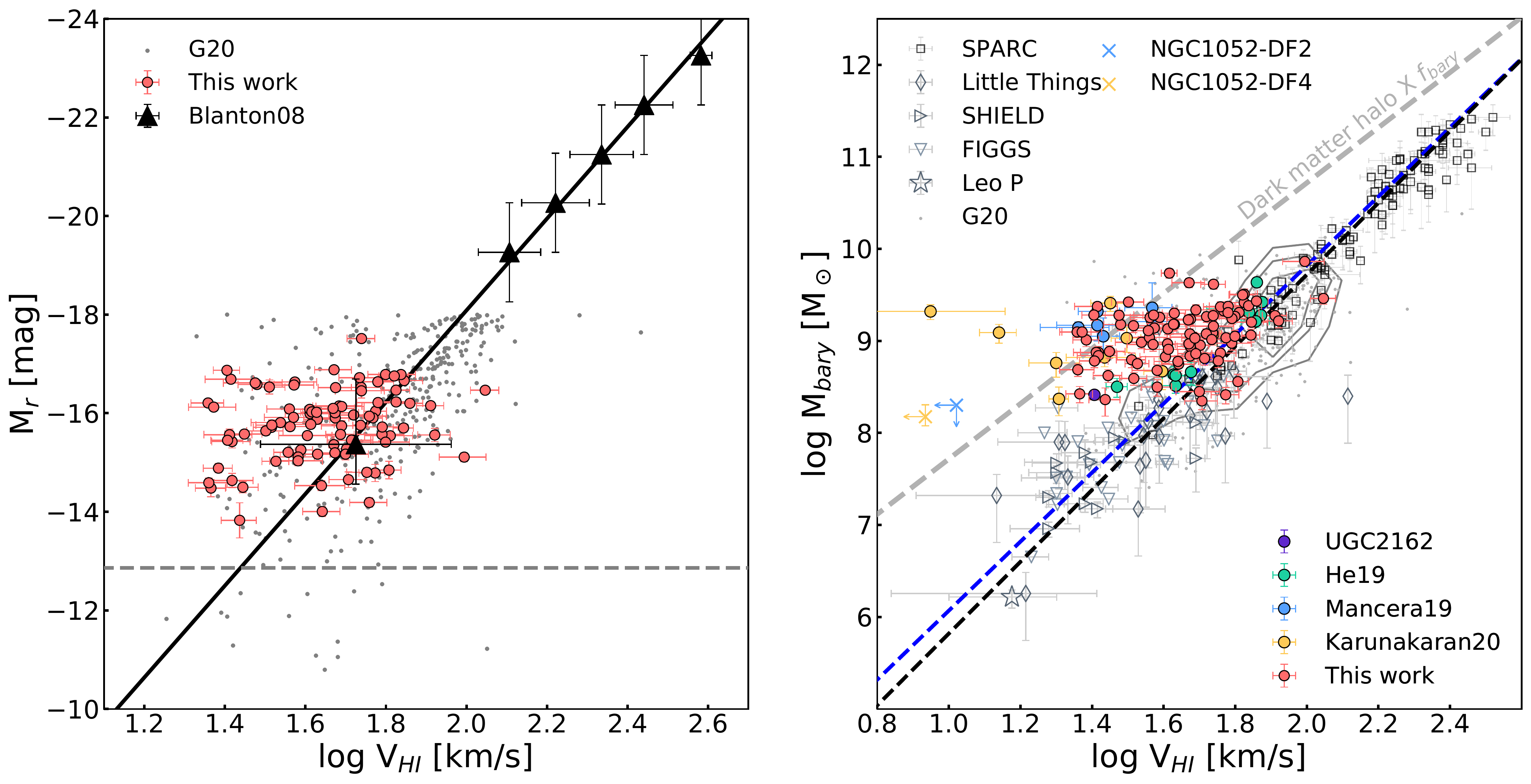}
\caption{{\bf Left: Tully-Fisher Relation}: {\it r}-band absolute magnitude vs. circular velocity. Red-filled circles are 88 H{\sc i}-rich ultra-diffuse galaxies (HUDGs) in this work. Dark triangles with error bars present massive galaxies from \cite{2008ApJ...682..861B} with the least-square fitting shown as the black straight line. Grey dots are dwarf galaxies from $\alpha.40$ (\citetalias{2020NatAs...4..246G}). The light grey line shows the incompleteness limit (see text for more details). 
{\bf Right: Baryonic Tully-Fisher Relation}: baryonic mass vs. circular velocity. Normal galaxies are shown with grey symbols, and HUDGs are shown with color-filled circles. The grey dots are \citetalias{2020NatAs...4..246G} dwarf galaxies with number density contours shown in grey lines. Grey diamonds, grey right triangles, and grey down-triangles are taken from Little Things \citep{2015AJ....149..180O} and SHIELD \citep{2016ApJ...832...89M} and FIGGS \citep{2008MNRAS.386..138B}. The grey star is Leo P \citep{2015ApJ...812..158M}. Blue \citep[NGC1052-DF2,][]{2018Natur.555..629V}, and yellow \citep[NGC1052-DF4,][]{2019ApJ...874L...5V} crosses are the dark matter deficient UDGs with the upper limits shown with the arrows. Green-filled circles are HUDGs taken from ALFALFA $\alpha.40$ \citep{2019ApJ...880...30H}, blue-filled circles are HUDGs using VLA \citep{2019ApJ...883L..33M}, yellow-filled circles are HUDGs observed with GBT \citep{2020ApJ...902...39K} and the blue-filled circle is the UGC 2162 \citep{2019MNRAS.488.3222S}. Grey hollow squares are for SPARC massive galaxies \citep{2016ApJ...816L..14L} for which the least-square fitting is presented with the black dashed line. The blue dashed line is the best linear fit for edge-on gas-dominated ALFA galaxies \citep{2016A&A...593A..39P}. Grey dashed line indicated universal ($f_{bary}=0.157$) baryonic mass within $M_{200}$ \citep{2012AJ....143...40M}.  
\label{fig:TFbtf}}
\end{figure*}

\section{Results} \label{sec:res}
In this section, we present the TFR and the BTFR of the HUDGs and compare them with those of normal dwarf galaxies and massive galaxies. We also show predictions by hydrodynamical cosmological simulation for further inference. 

\subsection{Tully-Fisher and Baryonic Tully-Fisher Relations} \label{sec:tfr}
We show the optical TFR of HUDGs in the left panel of Fig. \ref{fig:TFbtf}, in comparison with massive isolated galaxies in SDSS \citep{2008ApJ...682..861B}. It shows that the normal dwarf galaxies follow the TFR determined by massive spiral galaxies (massive-TFR). This has also been found by \cite{2008MNRAS.386..138B} using the Faint Irregular Galaxy GMRT survey (FIGGS) and by \cite{2017AJ....153....6K} in the Local Volume using the Updated Nearby Galaxy Catalog (UNGC). 

HUDGs flatten out towards low circular velocities in the TFR diagram. Those with $logV_{HI} > 1.6$ [km s$^{-1}]$ fall on the massive-TFR, while those with $logV_{HI} < 1.6$ [km s$^{-1}]$ deviate from the massive-TFR towards higher luminosity. At the lowest circular velocity, the deviation could reach as high as $3\, \sigma$ ($1\, \sigma$ is determined with $logV_{HI}>1.6$ dwarf galaxies). Similar distributions are also found in their stellar mass TFR (see Fig. \ref{fig:smTFR} in the Appendix).

The right panel of Fig. \ref{fig:TFbtf} presents the BTFR of HUDGs in comparison with massive spiral galaxies and normal dwarf galaxies. Massive spiral galaxies include 118 galaxies from Spitzer Photometry and Accurate Rotation Curves \citep[SPARC;][]{2016ApJ...816L..14L} with extended H{\sc i} rotation curves and Spitzer photometry at 3.6 $\mu$m, and 97 edge-on gas-dominated massive galaxies from ALFA catalogue \citep{2016A&A...593A..39P}. Dwarf galaxies include 26 nearby dwarf galaxies from Little Things \citep{2015AJ....149..180O}, 12 dwarf galaxies from Survey of H{\sc i} in Extremely Low-mass Dwarfs (SHIELD; \citealt{2016ApJ...832...89M}, VLA), the Leo P \citep{2015ApJ...812..158M}, and Local Volume dwarf galaxies from FIGGS \citep[GMRT;][]{2008MNRAS.386..138B}. \citetalias{2020NatAs...4..246G} dwarf galaxies (324) from $\alpha.40$ are presented as grey dots, with contours indicating the number density. Similar to the TFR, normal dwarf galaxies tend to follow the BTFR determined by massive galaxies (massive-BTFR) though with larger scatters than massive galaxies. This finding is consistent with previous work \citep{2008MNRAS.386..138B,2017AJ....153....6K}.

HUDGs also flatten out towards low circular velocity in the BTFR diagram. Different from the TFR, HUDGs are mostly above the median value of the BTFR of normal dwarf galaxies. The significance of the deviation is a strong decreasing function of the circular velocity. HUDGs with circular velocity $logV_{HI} > 1.6$ mostly lie on the upper 1 $\sigma$ region of normal dwarf galaxies, while HUDGs with $logV_{HI} < 1.6$ have much higher baryonic mass compared to normal dwarf galaxies. Quantitatively, using the scatter of dwarf galaxies with $1.6 < logV_{HI} < 2.1$ where the statistics are better, the mean deviation is 1.16 $\sigma$ at $V_{HI} > 40$ km s$^{-1}$, while the maximum deviation could reach $4.74\, \sigma$ at 26 km s$^{-1}$. Note that the significance of deviation could be an overestimation because the scatter at lower circular velocities could be even larger. 

UDGs with high baryon fractions have been reported in the literature. \cite{2018Natur.555..629V,2019ApJ...874L...5V} found two satellite UDGs deficient in dark matter using the dynamics of surrounding globular clusters. \cite{2019ApJ...880...30H} reported eleven edge-on HUDGs in $\alpha.40$ that are above the massive-BTFR. The isolated HUDG, UGC 2162, is above the $2.49\, \sigma$ level of BTFR from the normal dwarf galaxies \citep{2019MNRAS.488.3222S}. \cite{2019ApJ...883L..33M} found six HUDGs that deviate from the massive-BTFR significantly using VLA data. \cite{2020ApJ...902...39K} found nine gas-rich HUDGs deviate from the massive-BTFR using GBT data. All these reported HUDGs lie within the region defined by our HUDGs, which suggests that the UDGs in the literature do not constitute a distinct class of objects above the BTFR of normal dwarf galaxies, but are more probably part of a wide distribution of objects scattered toward the low-velocity side of the BTFR. 

\begin{figure}[ht!]
\plotone{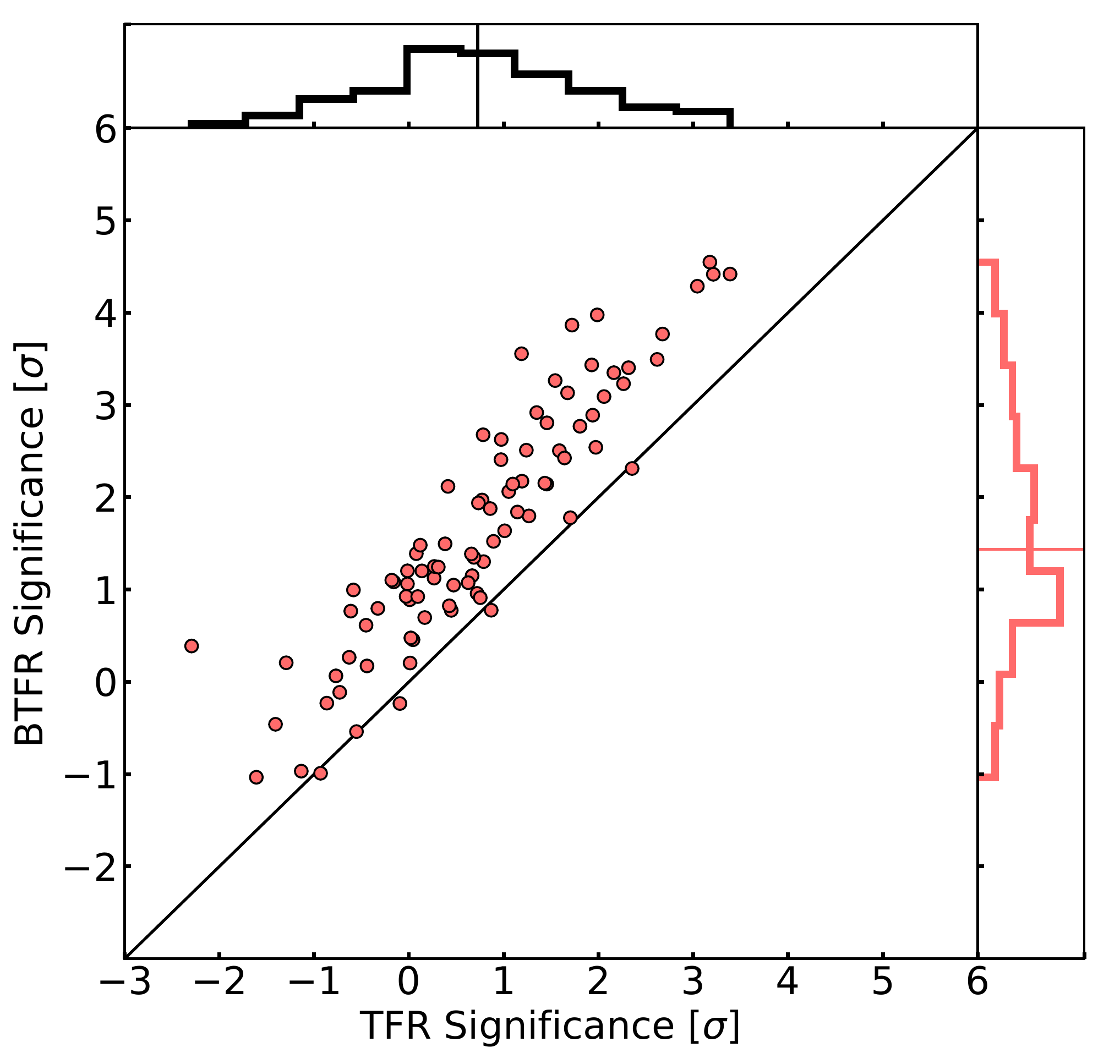}
\caption{{\bf Significance of the deviation of H{\bf \sc i}-rich ultra-diffuse galaxies}: deviation significance in BTFR vs. in TFR. The significance histograms of TFR (black) and BTFR (red) are shown on the upper and right sides of the figure, with the median values as shown with black/red lines. The significance is calculated according to the scatter of the \citetalias{2020NatAs...4..246G} dwarf with $1.6 < logV_{HI} < 2.1$ in the TFR and BTFR, respectively.
\label{fig:sig}}
\end{figure}

Fig. \ref{fig:TFbtf} also shows that HUDGs' deviation from massive-BTFR is more significant than their deviation from the massive-TFR. We quantify the deviation from TFR and BTFR according to the scatter of dwarf galaxies (\citetalias{2020NatAs...4..246G}) with $1.6 < logV_{HI} < 2.1$, respectively. Fig. \ref{fig:sig} shows clearly that the significance is systemically larger for the BTFR than that for the TFR by about 1 $\sigma$. The slope is larger than 1, i.e. more prominent at larger significance. In other words, the difference is more significant towards lower circular velocities. 

\subsection{What determines the difference between HUDGs' TFR and BTFR?}
\begin{figure}[ht!]
\plotone{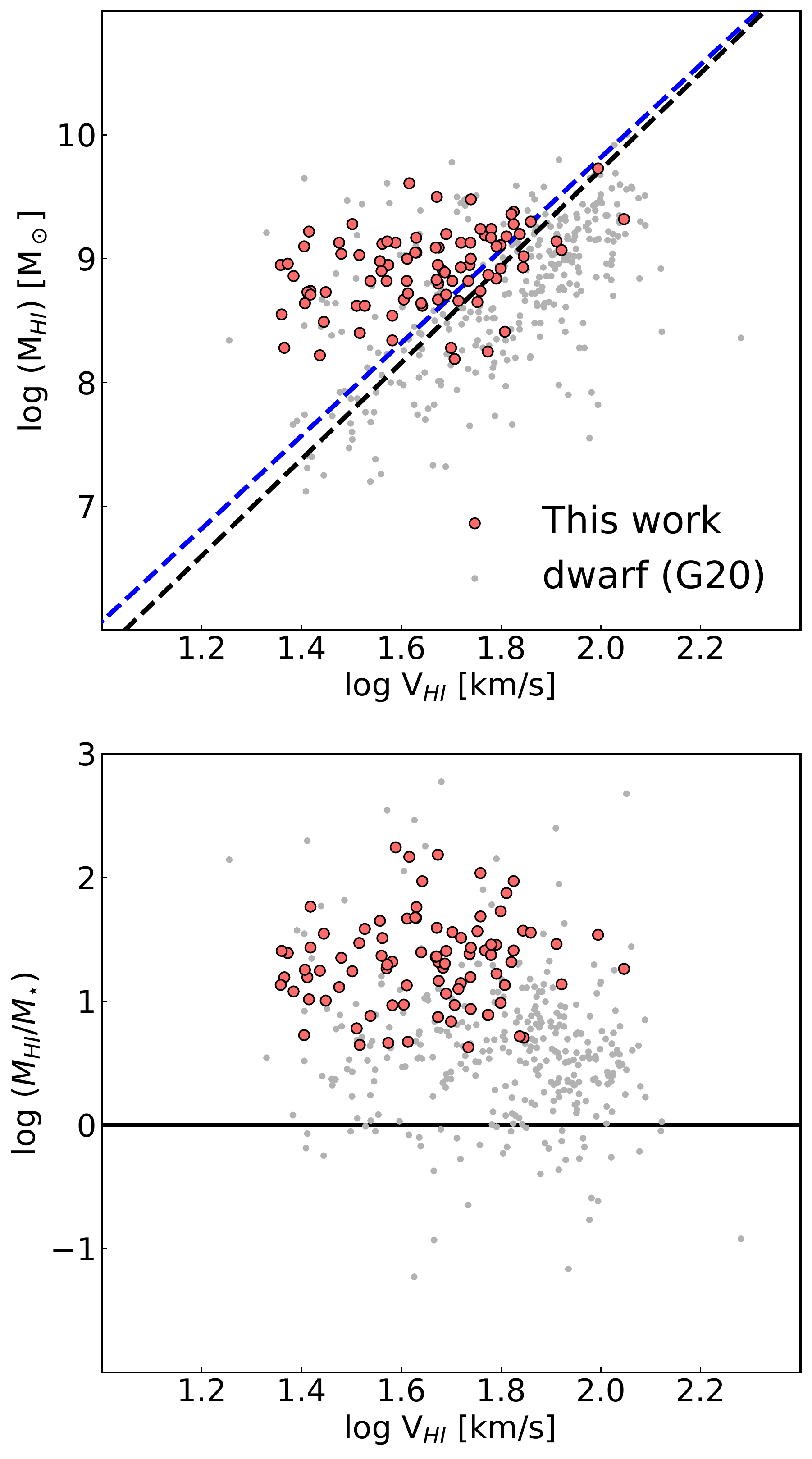}
\caption{{\bf Upper: H{\bf \sc i} mass vs. circular velocity.} The blue and black dashed lines are the best linear fit for massive galaxies with the same color coding as those in Fig. \ref{fig:TFbtf}. Grey dots are dwarf galaxies from \citetalias{2020NatAs...4..246G}. Red-filled circles are our 88 H{\sc i}-rich ultra-diffuse galaxies (HUDGs). {\bf Lower: Gas fraction vs. circular velocity}. The horizontal line denotes the 1:1 ratio between H{\sc i} mass and stellar mass. In general HUDGs have higher gas fraction than normal dwarf galaxies.
\label{fig:alfHI}}
\end{figure}

Why is the difference between HUDGs and normal dwarf galaxies smaller in the TFR than that in the BTFR? An intuitive explanation is that the HUDGs in our sample might have higher gas fractions. We test this hypothesis by comparing the H{\sc i} mass vs. circular velocity relation between HUDGs and normal dwarf galaxies in the upper panel of Fig. \ref{fig:alfHI}. It shows that H{\sc i} mass in HUDGs is generally higher than that in dwarf galaxies. For HUDGs the minimum H{\sc i} gas mass is $\sim 10^8M_{\odot}$, while for normal dwarf galaxies, it could be one order of magnitude lower. This is also reflected in the gas fraction vs. circular velocity relation in the lower panel. HUDGs have a median H{\sc i}-to-stellar mass ratio of 20.8 and a minimum H{\sc i}-to-stellar mass ratio of 4.3. For normal dwarf galaxies, the median and the minimum values are 3.72 and 0.06, respectively, much lower than that of the HUDGs. The high gas fraction of HUDGs has also been reported in \cite{2019MNRAS.490..566J} (see their Fig. 4). We thus expect that the high gas fraction could play an important role in pushing HUDGs above the BTFR of normal galaxies.

\begin{figure*}[ht!]
\plotone{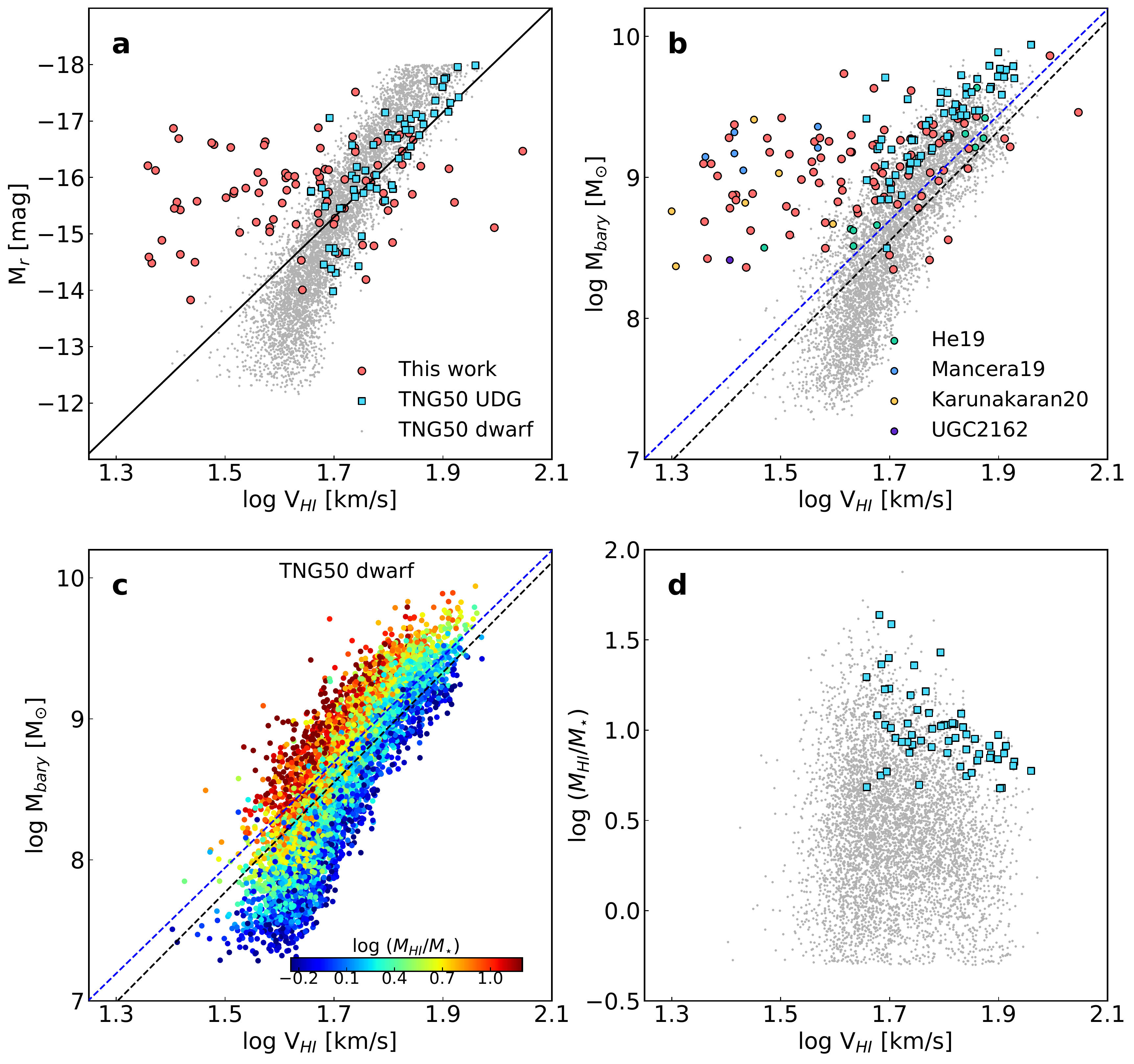}
\caption{{\bf Panel a and b: Tully-Fisher and baryonic Tully-Fisher relations in TNG50 simulation.} Grey dots and blue squares are isolated H{\sc i}-rich dwarf galaxies and H{\sc i}-rich ultra-diffuse galaxies (HUDGs) in the TNG50 simulation. Red circles are HUDGs in $\alpha.100$. The black line in panel a is the best fit of massive galaxies (same as the left panel of Fig. \ref{fig:TFbtf}). Black and blue dashed lines in panel b are the fittings of massive galaxies from observations (same color coding as the right panel of Fig. \ref{fig:TFbtf}). {\bf Panel c: gas fraction dependence of BTFR.} Dots are dwarf galaxies from TNG50. Different colors present galaxies with different gas fractions, as denoted by the color bar in the right bottom corner. {\bf Panel d: gas fraction vs. circular velocity.} HUDGs in TNG50 have higher gas fractions by selection. 
\label{fig:TNG50}}
\end{figure*}

We use Illustris-TNG simulation project\footnote{https://www.tng-project.org}, one of the most up-to-date cosmological hydrodynamical simulation sets, because of its completeness in statistics. We select dwarf galaxies from TNG50 \citep{2019MNRAS.490.3196P,2019MNRAS.490.3234N}, one of the high-resolution cosmological hydro-dynamical simulations. TNG50 traces $2\times2160^3$ dark-matter particles and gas cells in a period box of 51.7 Mpc on each side. The mass is $4.5\times10^5 M_\odot$ and $8.5\times10^5 M_\odot$ for each dark matter and baryon particle, respectively. Stellar masses were calculated using all star particles within each subhalo. We assume all cold hydrogen are H{\sc i}. We use the `Neutral Hydrogen Abundance' from the catalogue for H{\sc i} in non-star-forming cells, and calculate the H{\sc i} mass using the two-phase ISM model \citep[a modification of][]{2003MNRAS.339..289S} in star-forming cells. Instead of direct measurement of the H{\sc i} spectrum, we use the circular velocity at radii containing $90\%$ H{\sc i} mass \citep[see also ][]{2017MNRAS.464.2419S}. We select galaxies with at least 200 gas particles and 200 star particles to have a reliable estimation of the corresponding physical properties. Since most ALFALFA galaxies are in the fields, we focus on central dwarf galaxies in simulations with $M_r>-18$ and gas fraction larger than 0.5. Most of the simulated dwarf galaxies have circular velocities above $logV_{HI} > 1.6$. To mimic the HUDG selection in observations, we further require the simulated HUDGs to satisfy the following criteria: 1) the half stellar mass radius is larger than 1.5 kpc; 2) the mean surface brightness in the half mass radius is fainter than 24 mag arcsec$^{-2}$; and 3) the gas fraction is greater than 4.68 (95\% of observed HUDGs have gas fraction greater than this value).
 
We find a discrepancy between simulation and observations at low masses (Fig. \ref{fig:TNG50} a and b). Both TFR and BTFR bend down at around $10^9M_{\odot}$ in the simulation. Such discrepancy is consistent with what was found in previous studies \citep{2011ApJ...742...16T,2017MNRAS.464.2419S}.

Despite the discrepancy, the relative positions of HUDGs compared to normal dwarf galaxies are similar to what we find in the observations of both TFR and BTFR diagrams. HUDGs follow the same TFR as normal dwarf galaxies, while they mostly lie above the median value of the BTFR determined by normal dwarf galaxies. The deviation is mostly within the 1 $\sigma$ dispersion range of normal dwarf galaxies. 

This similarity allows us to use the simulation to answer why HUDGs are above the BTFR of normal galaxies. Panel c in Fig. \ref{fig:TNG50} shows the gas fraction dependence of the BTFR of dwarf galaxies (we find there is no gas fraction dependence in the TFR diagram). It shows that the higher the gas fraction, the higher the position in the BTFR diagram. In combination with panel d, which shows clearly a high gas fraction of the HUDGs, we conclude that the high gas fraction determines the higher baryonic mass of HUDGs in the BTFR diagram.

In summary, HUDGs flatten out towards low circular velocities both in TFR and BTFR diagrams. HUDGs with higher circular velocities follow the massive-TFR but deviate at low velocities. On the other hand, HUDGs lie above the massive-BTFR at all velocity ranges. The deviation is more significant in BTFR compared to that in TFR. The high gas fraction plays an important role in explaining this difference.

\subsection{Systemic uncertainties}
The detection limits of both optical (SDSS) and radio (ALFALFA) bands may bring additional systematic uncertainties to our study. By definition, HUDGs are more extended than typical dwarf galaxies and thus could have a higher luminosity limit for a complete sample. \cite{2004AJ....127..704K} reported the 3 $\sigma$ surface brightness isophote limit of the SDSS in {\it r}-band is 26.2 mag arcsec$^{-2}$. For our most face-on HUDGs ($b/a=0.7$) with the smallest size, 1.5 kpc, the luminosity limit is $M_r=-12.86$ (See Fig. \ref{fig:reTF} in Appendix). Ranking by optical size, luminosity limits at $50\%$ ($r_e=2.89$ kpc) and $90\%$ ($r_e=4.67$ kpc) are $M_r=-14.28$ and $M_r=-15.33$, respectively. For HUDGs smaller than $r_e=2.89$ kpc, the selection effects are not significant, while for those with larger sizes, the selection effects cannot be ignored. Moreover, the sensitivity limit of the ALFALFA survey would bias the sample to gas-rich UDGs. The deviation from normal TFR and BTFR might only hold for gas-rich UDGs, and their flat feature reflects the detection limit of ALFALFA. In other words, we cannot confirm the low dependency of the gas fraction in BTFR that shows in the high-mass galaxies still hold for low-mass galaxies ($M_{bary}<10^8M_\odot$). 

Besides, the uncertainty of intrinsic H{\sc i} velocity could be introduced by the inclination determination errors using the galaxies' optical morphology. The inclination parameter could be affected by non-axial symmetry and misalignment between the optical and H{\sc i} velocity field. Thus the value could be over- or under-estimated. The inclination correction derived from the b/a ratio is not likely affected by the biases from line width since there is no noticeable dependency (see Fig. \ref{fig:widthba}). Thus we consider the inclination correction can correct the projection effect on $w_{20}$ in general. Therefore, we do not expect that the systematic deviation of our HUDGs from the established TFR relation be caused by uncertainties in the inclination correction. However, the uncertainties in inclination correction may induce additional scatter to the distribution of the corrected $w_{20}$ in a random sense. We thus expect that the uncertainties would only increase the scatter but less likely cause systematic deviation.

The deviation from massive-TFR and massive-BTFR is more significant at low circular velocities. If their corrections were underestimated, the true higher circular velocities would push them toward the massive-TFR and massive-BTFR. As a consequence, the significance of the deviation could be reduced, especially at low circular velocities. 

We note that both the simulated HUDGs and those reported in the literature reside in the region covered by our 88 HUDGs in the BTFR diagram, suggesting that it could be a true feature of our HUDGs sample. 

\section{Discussion} \label{sec:dis}
The circular velocity is an indicator of the dark matter potential and the total mass, i.e. $M\sim V^{3}$. The positions of HUDGs in the BTFR diagram illustrate that HUDGs have higher baryonic mass compared to normal dwarf galaxies at a given circular velocity. This suggests inefficient feedback in HUDGs which could lead to more baryons in low-potential systems. This weak feedback scenario has also been reported by \cite{2019ApJ...883L..33M,2020MNRAS.495.3636M} and \cite{2019A&A...630A.140R}.

Analogs of the high BTFR-HUDGs could be the two dark matter deficient UDGs reported by \cite{2018Natur.555..629V,2019ApJ...874L...5V}. Note that pure rotation-supported disk galaxies are usually unstable under non-axisymmetric perturbations, which could lead to bar formation and heat up stellar systems \citep[e.g. ][]{1973ApJ...186..467O}. This has also been proven in recent simulations. For example, \cite{2022MNRAS.514.4008S} studied the UDG, AGC 114905 reported by \cite{2022MNRAS.512.3230M}, and found that the galaxy in the dark-matter-free model is unstable and disrupts in a rotation time. Despite the high baryon fraction, most of our 88 HUDGs are not dark-matter-free. In addition, random motions from stars and thermal motion from gas also help stabilize the systems. 

\cite{2019MNRAS.488.3298J} found that the dark matter deficient systems could be formed by tidal stripping in clusters. Dark matter is less condensed compared to baryons and could be stripped by tidal forces. Meanwhile, baryons could resist tidal stripping. As a consequence, more baryons are left in the system which pushes their BTFR above the normal dwarf galaxies. our 88 HUDGs are mostly found in fields where the environmental impact is minimized. Their formation might be similar to the dark matter deficient dwarf galaxies found in \citetalias{2020NatAs...4..246G}. Among the 88 HUDGs, 20 of them can be classified as isolated baryon-dominated dwarf galaxies. The formation of them cannot be reproduced in current cosmological simulations. These baryon-dominated dwarf galaxies could be game-changer laboratories in testing cosmology models and galaxy formation models. 

\section{Summary} \label{sec:sum}
We study Tully-Fisher and baryonic Tully-Fisher relations of 88 H{\sc i}-rich UDGs (HUDGs) selected from ALFALFA. It is the largest sample of HUDGs with dynamical information. We compare the HUDGs' TFR and BTFR with those of normal dwarf galaxies and explore their origins. The key results are listed as follows. 
 
1) Dwarf galaxies follow the TFR and the BTFR determined by massive galaxies. 

2) HUDGs flatten out towards low circular velocities in the TFR diagram. They follow the TFR determined by normal dwarf galaxies at $V_{HI} > 40$ km s$^{-1}$ but deviate towards higher luminosity at $V_{HI} < 40$ km s$^{-1}$. 

3) HUDGs flatten out towards low circular velocities in the BTFR diagram. Different from TFR, most HUDGs are above the median BTFR defined by normal dwarf galaxies. The deviation is a decreasing function of circular velocity. At circular velocity $V_{HI} > 40$ km s$^{-1}$, HUDGs mostly lie above normal dwarf galaxies with a mean deviation of $1.16\, \sigma$, while at circular velocity $V_{HI} < 40$ km s$^{-1}$ the deviation could reach as high as $4.74\, \sigma$. HUDGs reported in the literature all reside in the region defined by our HUDGs.

4) HUDGs' deviation from massive-BTFR is systematically more significant than that from massive-TFR. 

5) The selection-induced high gas fraction could play an important role in explaining their high positions in the TFR and BTFR. It is also supported by modern simulations.

In general, the conclusion still could be affected by the selection effects due to the depths of optical and H{\sc i} surveys, the inclination corrections, and systemic asymmetries. Further surveys with deeper photometries and well-resolved H{\sc i} velocity fields would help to reach more conclusive results.

\acknowledgments
We thank the anonymous referee for the constructive comments. This work is supported by the National Key Research and Development of China (No.2022SKA0110201, 2018YFA0404503), the National Natural Science Foundation of China (NSFC) grants (No.12033008, 11622325, 11988101), the K.C.Wong Education Foundation, and the science research grants from the China Manned Space Project (CMSP) with NO.CMS-CSST-2021-A03 and NO.CMS-CSST-2021-A07. Q.G. acknowledges support from the joint Sino-German DFG research Project ``The Cosmic Web and its impact on galaxy formation and alignment" (DFG-LI 2015/5-1, NSFC No.11861131006) and the support of the Shanghai International partners project (No.19590780200). Z.Z. is supported by NSFC grants (No.11988101, 12041302, 11703036, and U1931110), CAS Interdisciplinary Innovation Team (JCTD-2019-05), and No.CMS-CSST-2021-A08. C.-W.T. is supported by NSFC grant (No.12041302). H.Z. is supported by NSFC grants (No.12122303, 11973039), and CAS Pioneer Hundred Talents Program. ZY. Z. is supported by NSFC grants (No.12041305, 12173016), and CMSP grants (NO.CMS-CSST-2021-A08, NO.CMS-CSST-2021-A07). ZY. Z. also acknowledges the Program for Innovative Talents, Entrepreneur in Jiangsu.


\appendix
\renewcommand\thefigure{\Alph{section}}

\begin{deluxetable*}{ccccccccc}
\tablenum{A.1}
\tablecaption{Parameters of H{\sc i}-rich ultra-diffuse galaxies. \label{tab:HUDG}}
\tablewidth{0pt}
\tablehead{\colhead{AGCNr} & \colhead{Dist} & \colhead{$w_{20}$} & \colhead{$M_g$} & \colhead{$M_r$} & \colhead{log $M_{st}$} & \colhead{log $M_{HI}$} & \colhead{log $M_{bary}$} & \colhead{b/a} \\
 \colhead{AGC } & \colhead{Mpc} & \colhead{km s$^{-1}$} & \colhead{mag} & \colhead{mag} & \colhead{$M_{\odot}$} & \colhead{$M_{\odot}$} & \colhead{$M_{\odot}$} & \colhead{} }
\decimalcolnumbers
\startdata
{\bf 103435} & 28.5$\pm$2.3 & 39.23$\pm$1.0 & -14.25$\pm$0.18 & -14.48$\pm$0.18 & 7.09$\pm$0.37 & 8.28$\pm$0.09 & 8.42$\pm$0.09 & 0.56 \\
{\bf 105118} & 72.8$\pm$2.2 & 70.7$\pm$2.42 & -16.02$\pm$0.07 & -16.15$\pm$0.07 & 7.63$\pm$0.16 & 8.9$\pm$0.06 & 9.04$\pm$0.06 & 0.697 \\
{\bf 105466} & 77.0$\pm$2.4 & 46.32$\pm$3.36 & -14.23$\pm$0.08 & -14.89$\pm$0.09 & 7.78$\pm$0.2 & 8.86$\pm$0.07 & 9.01$\pm$0.07 & 0.348 \\
{\bf 102791} & 75.4$\pm$2.2 & 114.37$\pm$2.84 & -16.25$\pm$0.07 & -16.47$\pm$0.07 & 7.87$\pm$0.16 & 9.28$\pm$0.05 & 9.42$\pm$0.05 & 0.545 \\
{\bf 102269} & 74.8$\pm$2.1 & 110.22$\pm$3.21 & -16.15$\pm$0.06 & -16.66$\pm$0.06 & 8.31$\pm$0.15 & 9.02$\pm$0.06 & 9.2$\pm$0.06 & 0.637 \\
\enddata
\tablecomments{Col.(1): $\alpha.100$ ID number. Col.(2): distance. Col.(3): instrumental boarding corrected 20\% peak line width. Col.(4) and Col.(5): g-band and r-band absolute magnitude. Col.(6): stellar mass. Col.(7): H{\sc i} mass. Col.(8): total baryonic mass. Col.(9): g-band axis ratio.}
\end{deluxetable*}

\section{distance}
\begin{figure}[ht!]
\plotone{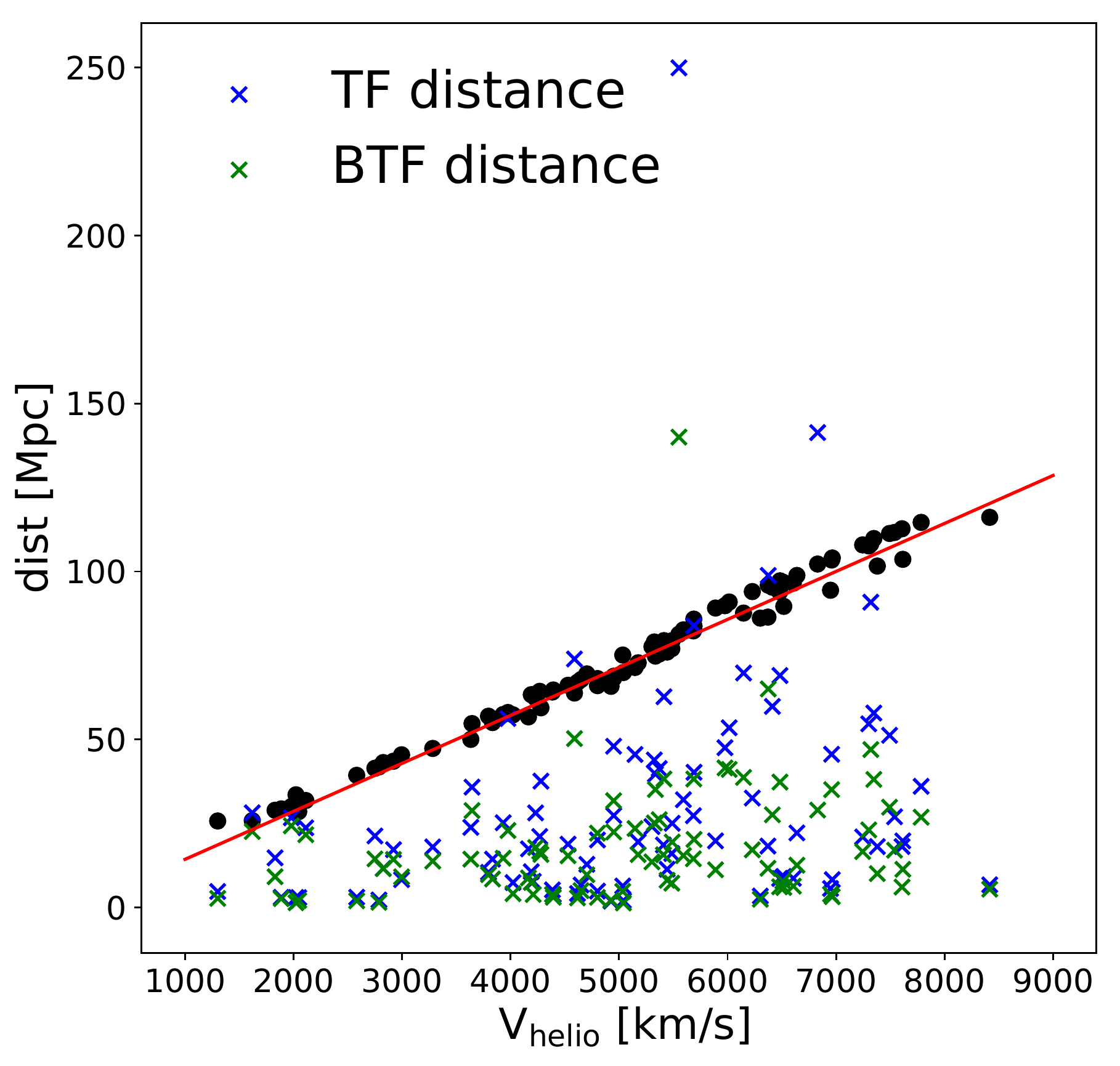}
\caption{{\bf Distance vs. heliocentric velocity for 88 H{\bf \sc i}-rich UDGs.} The blue (green) crosses are distances inferred from TFR (BTFR). Black dots are distances taken from the ALFALFA catalogue, and the red line indicates the Hubble flow distance. 
\label{fig:dist}}
\end{figure}

 The ALFA team applies a primary literature distance and a secondary distance from the Tully-Fisher relation (TFR) at $cz<6000$ km s$^{-1}$. At $cz>6000$ km s$^{-1}$, they adopt the distance using the Hubble flow. If a galaxy's distance is given by the TFR, it is not suitable for the study of TFR and baryonic Tully-Fisher relation (BTFR). Here we investigate the fraction of H{\sc i}-rich UDGs (HUDGs) whose distances are given by the TFR. We show the distance vs. heliocentric velocity in Fig. \ref{fig:dist}. Black dots are the distance given by ALFA used in this work. Blue and green crosses are the distances inferred from TFR or BTFR (using $w_{50}$/2), respectively. It shows that our sources are consistent with Hubble law (red line) even at $cz<6000$ km s$^{-1}$, and the distances are mostly above the distances inferred by TFR and BTFR. Most of the distances of our HUDGs are not the distance inferred from TFR or BTFR. 
 
\section{$w_{20}$ in ALFALFA}
\begin{figure}[ht!]
\plotone{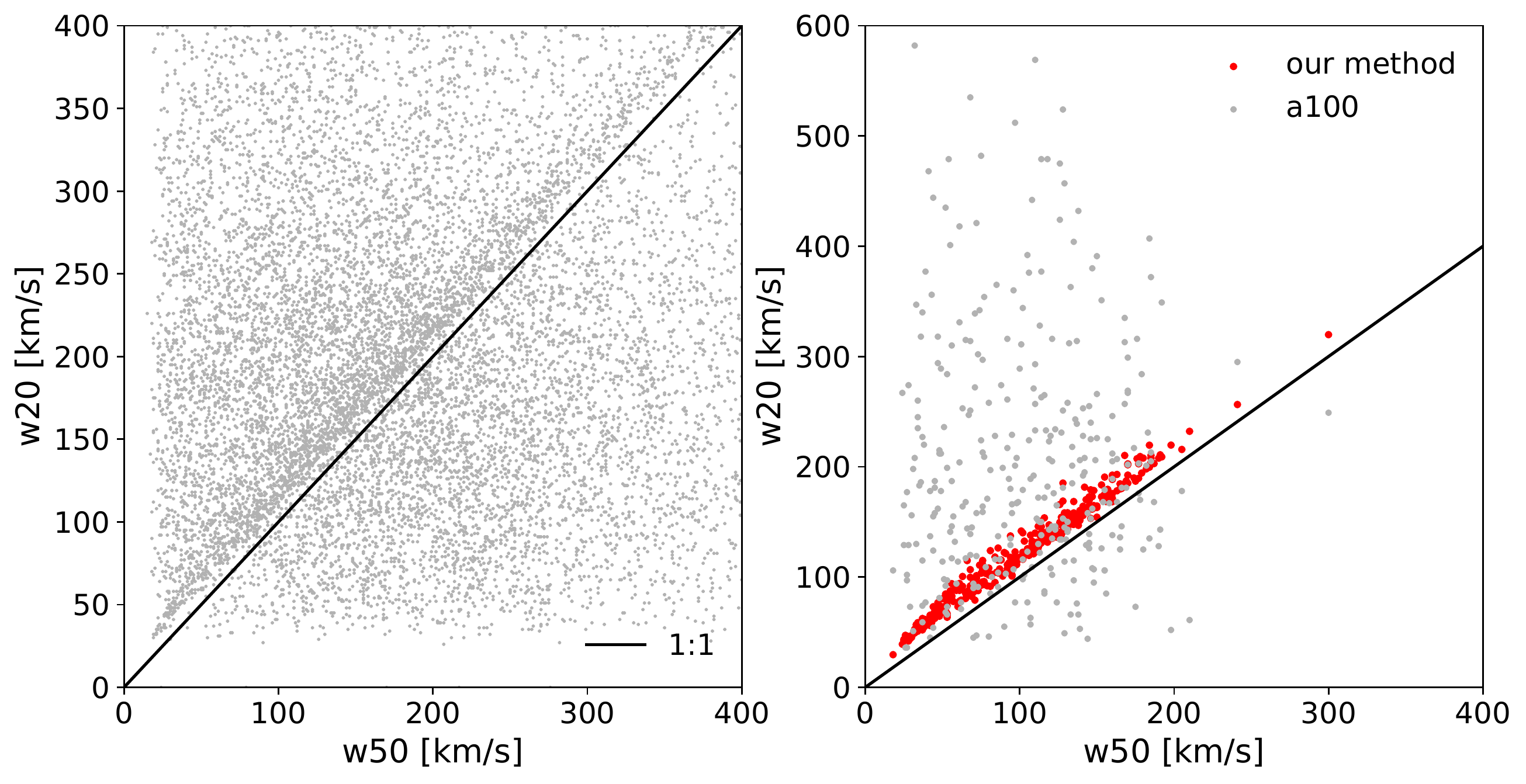}
\caption{{\bf $\bf w_{20}$ vs. $\bf w_{50}$. Left:} both values are taken from $\alpha.100$ {\bf Right:} $w_{50}$ are taken from $\alpha.100$, red dots are the re-processed values for our method.
\label{fig:w20alf}}
\end{figure}

In galaxy spectrum, $w_{20}$ and $w_{50}$ should have good correlation and usually $w_{20}$ is not less than $w_{50}$. We select sources with $SNR>10$, a high signal-to-noise ratio sample, and show the $w_{20}$ vs. $w_{50}$ both taken from the $\alpha.100$ catalogue in the left panel of Fig. \ref{fig:w20alf}. The correlation between the $w_{20}$ and $w_{50}$ is rather weak and the scatter is very big. Focusing on dwarf galaxies, we reprocess 324 dwarf galaxies (\citetalias{2020NatAs...4..246G}) with $SNR>10$, b/a: 0.3-0.6 and $M_r>-18$, and compare the $w_{20}$ and $w_{50}$ in $\alpha.100$ with the reprocessed ones in the right panel. Our measurements show a clear correlation between $w_{20}$ and $w_{50}$, whereas the scatter of $w_{20}$ from $\alpha.100$ is much larger than ours. In addition, the reprocessed $w_{20}$ are systematically larger than $w_{50}$, as expected. 

\section{stellar mass Tully-Fisher relation}
\begin{figure}[ht!]
\plotone{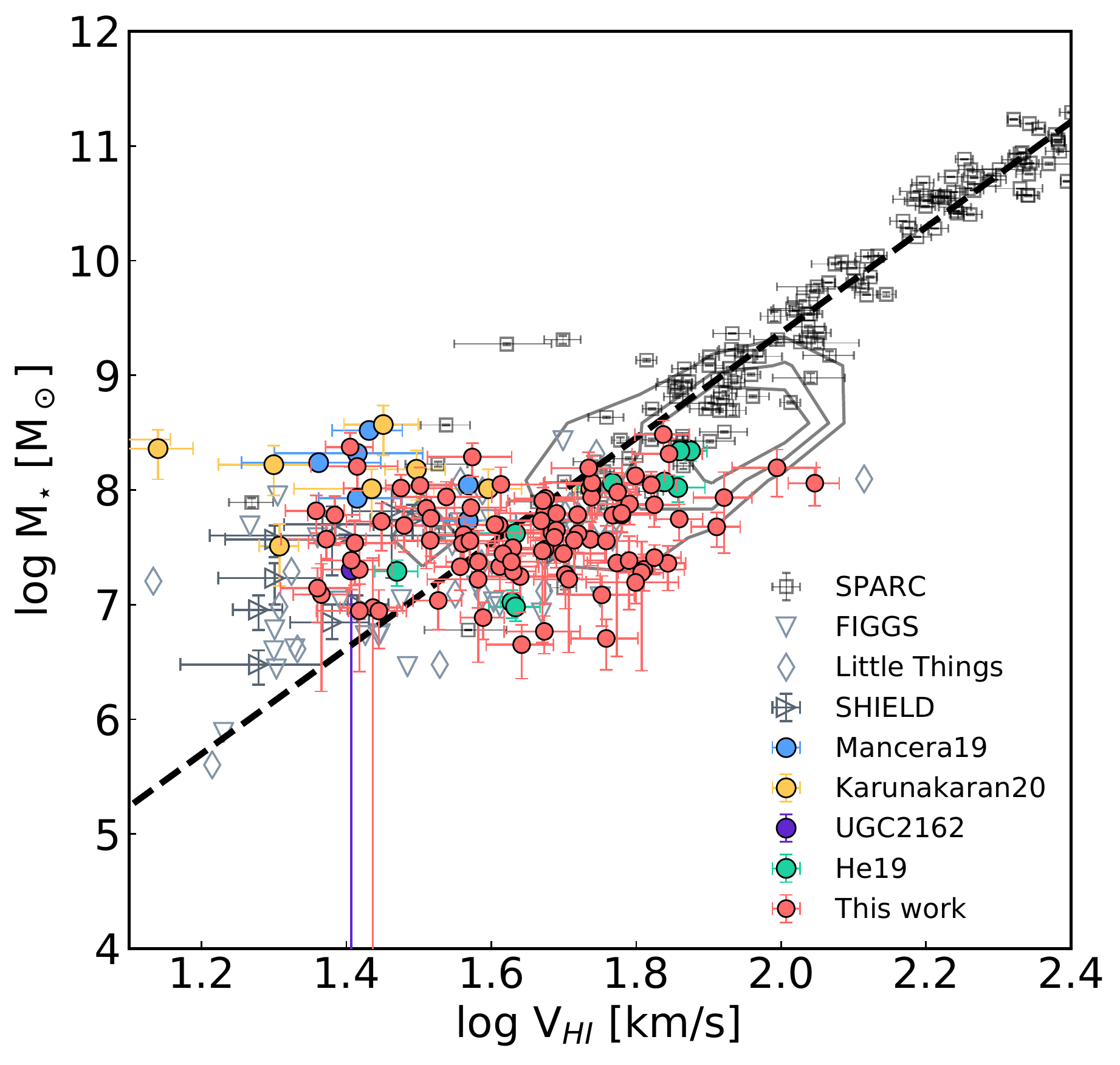}
\caption{ {\bf Stellar mass Tully-Fisher Relation:} symbols type and colors are the same coding as those in Fig. \ref{fig:TFbtf}. The dashed line denotes the best fit of the SPARC galaxies.
\label{fig:smTFR}}
\end{figure}

We show the stellar mass TFR (smTFR) in Fig. \ref{fig:smTFR}. The stellar mass is derived following Bell's algorithm (see detail in sec. \ref{sec:mass}). Note we do not correct for any dust extinction assuming that it can be ignored for dwarf galaxies. This could underestimate the derived stellar mass, which contributes to the offset between dwarf galaxies (grey contour) and massive galaxies. Same to the luminosity TFR, HUDGs flatten out in the smTFR diagram. Those with $V_{HI}>40$ km s$^{-1}$ follow the smTFR of normal dwarf galaxies, while those with lower circular velocities are far above the smTFR of normal galaxies. Interestingly, non-UDGs with low circular velocities, including those detected in Little Things and SHIELD, also lie above the smTFR, reflecting the selection effects discussed in the main text. 

\section{size distribution in the Tully-Fisher relation diagram}
\begin{figure}[ht!]
\plotone{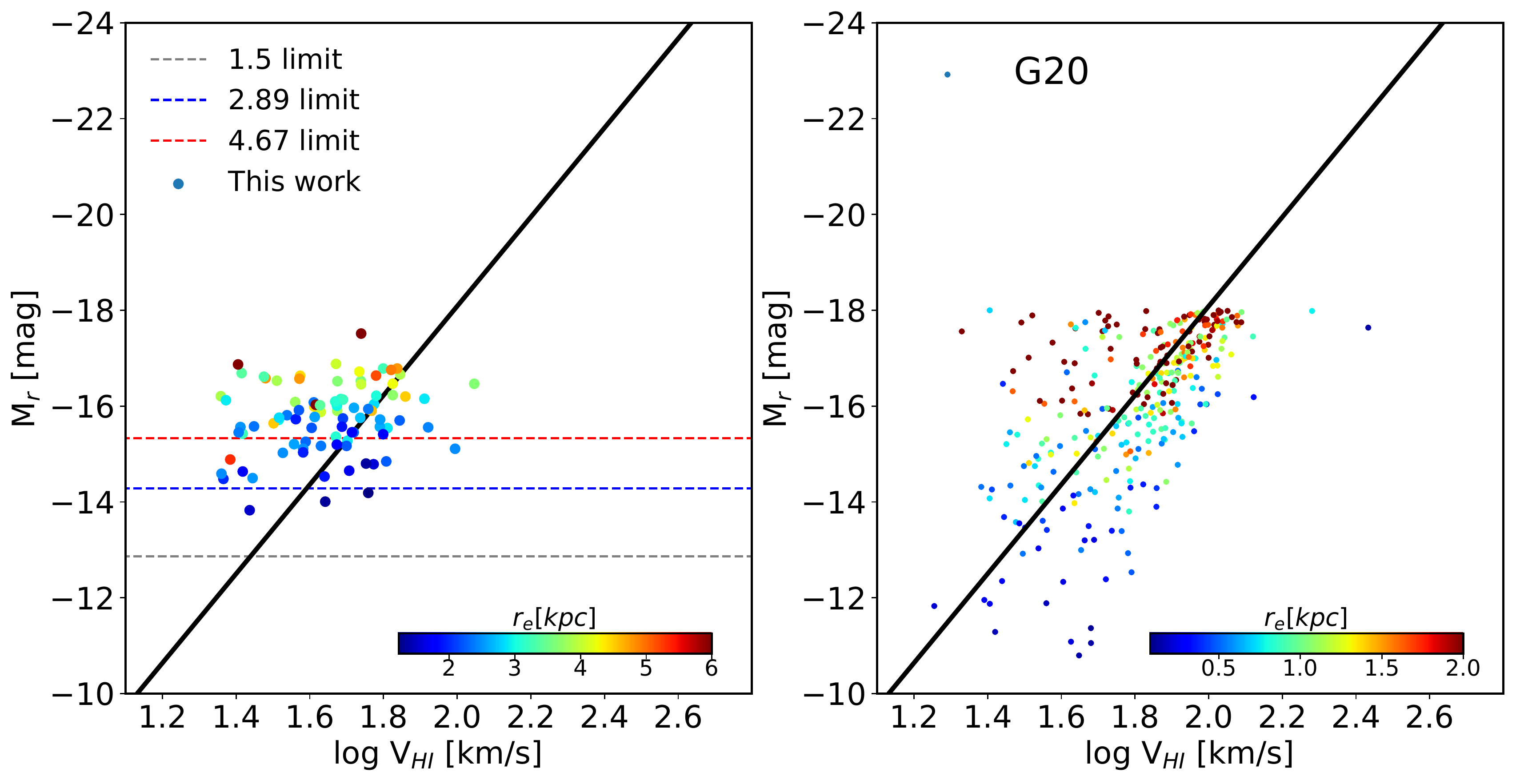}
\caption{ {\bf Half-light radius in Tully-Fisher relation diagram:} {\bf Left}: HUDGs, {\bf Right}: \citetalias{2020NatAs...4..246G} dwarf galaxies. Colored as half-light radii ($r_e$), the grey/blue/red dashed lines indicated the luminosity limits of minimal/$50\%$/$90\%$ size of HUDGs by $3\, \sigma$ of background. 
\label{fig:reTF}}
\end{figure}

Considering the luminosity limit, the $3\, \sigma$ surface brightness isophote limit of the SDSS in the {\it r}-band is 26.2 mag arcsec$^{-2}$ \citep{2004AJ....127..704K}. The luminosity limit is a function of galaxy size at a fixed axis ratio. Based on the definition of UDGs, the minimal half-light radius is 1.5 kpc, which corresponds to a luminosity limit of $M_r=-12.86$ for our most face-on HUDGs ($b/a=0.7$). While, for the median size of our HUDGs ($r_e=2.89$ kpc), the luminosity limit is $M_r=-14.28$. We show the size distribution in the TFR in Fig. \ref{fig:reTF}, where the grey/blue/red dashed lines indicate the luminosity limits of minimal/$50\%$/$90\%$ size of HUDGs. Indeed for UDGs with larger sizes, they are below the detection limit resulting in a magnitude incompleteness issue. Next-generation surveys with deeper photometries are needed for more robust results.

\section{Velocity corrections uncertainties}
\begin{figure}[ht!]
\plotone{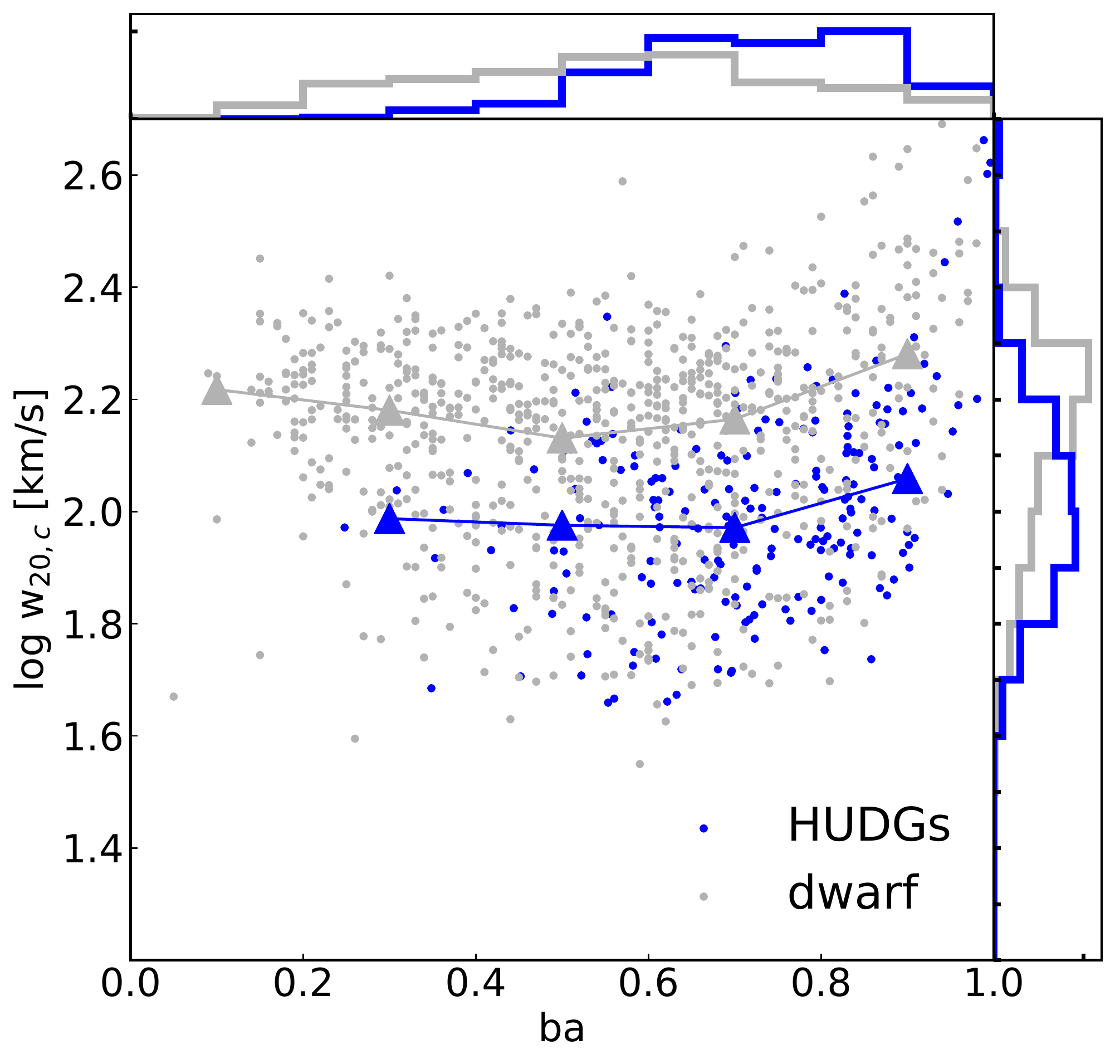}
\caption{{\bf Inclination-corrected velocity width vs. axis ratio (b/a)}. Grey and blue dots are for normal dwarf galaxies and HUDGs (without b/a cut), respectively. Curves represent the median values. The distribution of the b/a ratios and the inclination-corrected velocity width are presented on the top and right sides. The offset between normal dwarf galaxies and HUDGs is caused by their different mass distributions.  
\label{fig:widthba}}
\end{figure}

Without the information of the H{\sc i} velocity fields, we adopt optical inclination to correct for the true circular velocity. However, the optical inclination estimation could be affected by the non-axial symmetry and there could be a misalignment between optical inclination and H{\sc i}-velocity-field inclination. We test whether our treatment could lead to biased circular velocities. If there is systematic bias caused by the inclination correction, one would expect the inclination-corrected velocity varies with optical inclinations (b/a ratios). In Fig. \ref{fig:widthba}, we show the corrected velocity width vs. b/a ratio. Here we include HUDGs both with b/a $<$0.7 and b/a $>$ 0.7. Normal dwarf galaxies are selected from $\alpha.40$ with $M_r>-18$ and SNR of H{\sc i} spectrum greater than 10. It shows clearly that inclination-corrected velocity width is independent of the ba ratio both for normal dwarf galaxies and for HUDGs. The slight up-turn at $b/a>0.8$ could be due to an overestimate of the nearly face-on galaxies. We discard those with ba greater than 0.7 in the final HUDG sample. It suggests our simple treatment does not induce any biased correction systematically. 
The offset between the normal dwarf galaxies and HUDGs is due to the fact that dwarf galaxies samples are somewhat more massive than HUDGs.

\end{CJK*}
\end{document}